\documentclass[10pt,twocolumn,letterpaper]{article}

\usepackage[pagenumbers]{cvpr} 
\usepackage{multirow}
\usepackage{url}
\usepackage{amssymb}
\usepackage{pifont}
\usepackage{cuted}
\usepackage{amsmath}
\usepackage{graphicx}
\usepackage{booktabs}
\usepackage[table]{xcolor}
\usepackage{comment}
\usepackage{wrapfig}
\usepackage{caption}

\newcommand{\mS}{\mathcal{S}}
\newcommand{\mM}{\mathcal{M}}

\definecolor{cvprblue}{rgb}{0.21,0.49,0.74}
\usepackage[pagebackref,breaklinks,colorlinks,allcolors=cvprblue]{hyperref}

\def\paperID{2262} 
\def\confName{CVPR}
\def\confYear{2026}

\title{Human Geometry Distribution for 3D Animation Generation}

\author{
Xiangjun Tang \qquad Biao Zhang\qquad
Peter Wonka\thanks{Corresponding author.}\\
King Abdullah University of Science and Technology\\
{\tt\small \{xiangjun.tang, biao.zhang, peter.wonka\}@kaust.edu.sa}
}

\begin{document}
\maketitle

\begin{strip}
    \centering
    \vspace{-3em}
    \includegraphics[width=1.0\linewidth]{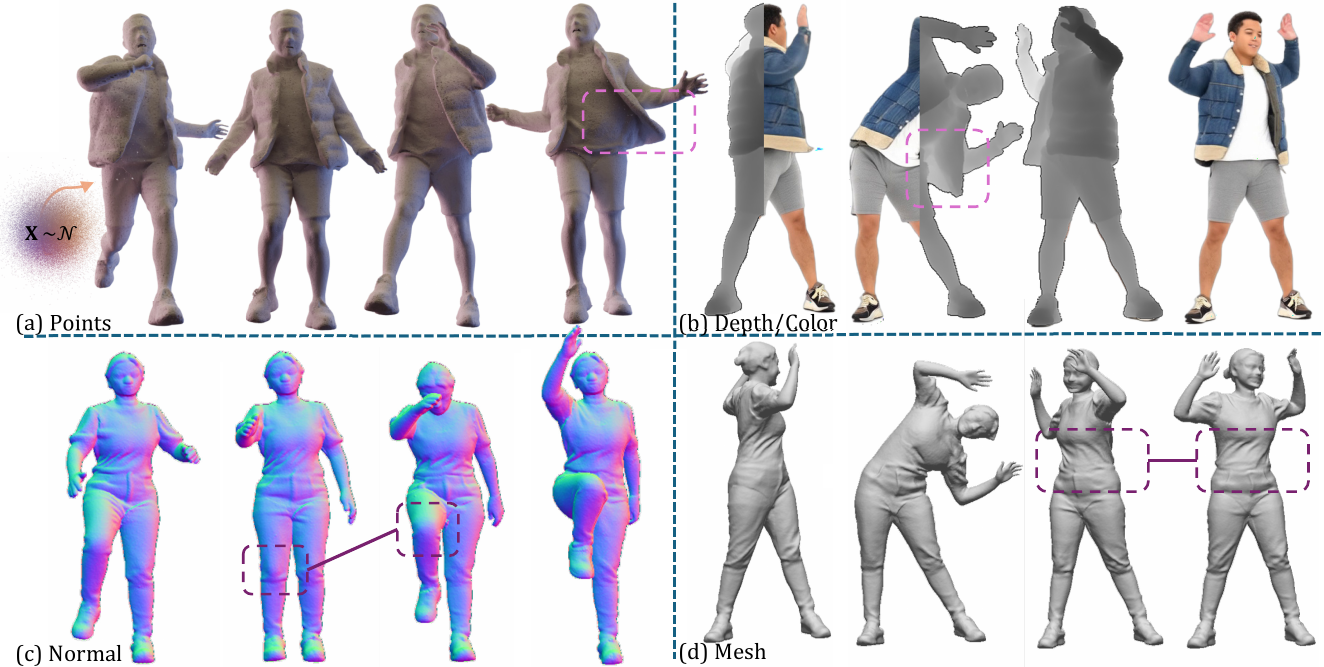}
    \vspace{-1em}
    \captionof{figure}{Our generative framework produces diverse avatar geometry sequences from noise, with geometries represented as points (a). For visualization, these points can be rendered via Gaussian splatting (GS), producing depth images (b) and normal images (c). Colors (b) can then be obtained by GS optimization, using a depth-guided video generation model (Wan 2.1), while the normal images (c) effectively highlight fine folds and wrinkles. Our synthesized geometries are of high quality and can be directly converted into meshes (d) via Poisson reconstruction. The highlighted regions demonstrate fine-grained garment dynamics that faithfully follow human motion.}
    \label{fig:teaser}
\end{strip}
\begin{abstract}
Generating realistic human geometry animations remains a challenging task, as it requires modeling natural clothing dynamics with fine-grained geometric details under limited data.
To address these challenges, we propose two novel designs. First, we propose a compact distribution-based latent representation that enables efficient and high-quality geometry generation. We improve upon previous work by establishing a more uniform mapping between SMPL and avatar geometries. Second, we introduce a generative animation model that fully exploits the diversity of limited motion data. We focus on short-term transitions while maintaining long-term consistency through an identity-conditioned design. These two designs formulate our method as a two-stage framework: the first stage learns a latent space, while the second learns to generate animations within this latent space.
We conducted experiments on both our latent space and animation model. We demonstrate that our latent space produces high-fidelity human geometry surpassing previous methods ($90\%$ lower Chamfer Dist.). The animation model synthesizes diverse animations with detailed and natural dynamics ($2.2 \times$ higher user study score), achieving the best results across all evaluation metrics.
\end{abstract}  
\section{Introduction}
Generating 3D human geometry animation is a fundamental task in visual generation and human modeling. The goal is to synthesize natural dynamics with fine-grained geometric details, which poses significant challenges.
First, capturing fine-grained details requires modeling subtle geometric structures such as folds and wrinkles. Second, learning natural dynamics is challenging due to the limited availability of 3D animation data, where models may easily overfit and fail to reproduce realistic garment deformation in response to human movement. 

Early methods~\cite{santesteban2019learning,santesteban2021self,patel2020tailornet} learn dynamics for specific garments, or model avatars from video or scanned data~\cite{xiang2021modeling,zhang2023closet,saito2021scanimate,ma2021scale,li2024animatable,zheng2024physavatar,wang2024pgahum,linLayGALayeredGaussian2024,lee2023dynamic,prokudin2023dynamic,habermann2021real,tiwari2021neural}.
While these approaches can synthesize plausible dynamics with limited data, they are not generative methods and fail to generalize to unseen avatars or garments.
In contrast, generative avatar models~\cite{corona2021smplicit,ma2020learning,zheng2024design2cloth,zakharkin2021point,laczko2024generative,zou2025generating,hong2022avatarclip} extend to diverse identities and offer better generalization, yet they struggle to preserve high-fidelity geometry and learn realistic clothing deformations. Overall, no existing approach satisfies both requirements.

To address these challenges, we propose two key designs. First, we propose a latent representation based on the Human Geometry Distribution (HuGeoDis)~\cite{tang2025generative}, which enables the synthesis of high-fidelity geometry from a compact latent representation. However, the original HuGeoDis suffers from imbalanced sampling: it requires a large number of points to adequately cover a geometry, and undersampled areas often lead to reconstruction artifacts. To mitigate this, we design a new training scheme that first establishes more uniform mappings between SMPL and avatar geometries, and then learns from these correspondences. This design enables high-quality geometry generation with significantly fewer points, thereby improving efficiency for long animation sequences. Second, we introduce a generative animation model that captures temporal dynamics from limited 3D human animation data. We employ a conditional diffusion model that models short-term transitions, which has been empirically shown to leverage diverse motion data more effectively than directly modeling long sequences~\cite{tang2022real}. Long sequences are generated auto-regressively from these transitions, with long-term consistency preserved via conditional inputs to the diffusion model.

We conduct experiments to validate both our distribution-based latent representation and the generative animation model. For the latent space, we evaluate reconstruction accuracy and efficiency, and further assess its performance for the static random avatar generation task, a standard benchmark in avatar generation. Our representation achieves higher reconstruction accuracy ($90\%$ lower Chamfer Distance) with greater efficiency and produces superior geometry quality compared to existing methods. For the generative animation model, experiments demonstrate its ability to synthesize diverse human geometries with detailed clothing dynamics that faithfully follow body motion, achieving $2.2\times$ higher user study score. 

Our contributions can be summarized as follows:
\begin{itemize}
    \item We present the first framework for generating 3D human geometry animations that captures natural dynamics with fine-grained geometric details.
    \item We propose a novel distribution-based latent space that is both compact and expressive, capable of efficiently representing high-fidelity human geometry.
    \item We develop a new generative animation model that synthesizes diverse 3D human animations even for a small 3D animation training dataset.
\end{itemize}

\section{Related work}
Our method is a generative approach that focuses on synthesizing detailed, pose-dependent human geometry with natural deformation. This connects it to work on garment deformation and avatar modeling, which investigate how motion affect clothing dynamics and surface details.
\subsection{Data-driven Garment Deformation}
Garment animation and simulation remains a long-standing and critical problem in computer graphics and vision. While physically based simulations yield high-fidelity results, they are hindered by high computational costs and poor scalability.
To overcome these limitations, data-driven approaches learn to approximate physically based deformations from examples of dressed characters with diverse poses and body shapes~\cite{santesteban2019learning}. For example, TailorNet~\cite{patel2020tailornet} represent clothing deformation as a function of human pose, shape, and garment style, enabling limited variation in garment geometry. To keep fine garment details and preserve correct collision between garment and body, GarNet~\cite{gundogdu2019garnet,gundogdu2020garnet++} incorporate curvature-based constraints and Santesteban \etal~\cite{santesteban2021self} introduce a self-supervised method to learn garment–body interactions. 
In parallel, skinning-based learning methods approximate dynamic garment motion under the linear blend skinning (LBS) formulation.
PBNS~\cite{bertiche2020pbns} learns a neural surrogate of physics-based simulation for tight-fitting garments, while Pan et al.~\cite{pan2022predicting} extend this idea to loose garments such as skirts by predicting bone-driven deformations.
DeePSD~\cite{bertiche2021deepsd} further predicts blend weights and transformations directly from garment meshes, achieving moderate generalization across unseen garments.
Although these methods can capture detailed garment geometry, they focus primarily on modeling clothing rather than the human body, are often tied to specific garment templates, and cannot support generative tasks that synthesize diverse garments.

\subsection{Animatable Avatar Modeling}
Animatable human modeling aims to reconstruct high-fidelity avatars that can be driven by novel poses, typically learned from multi-view videos or 3D scans.
Early approaches simplify this challenging task by using a predefined or reconstructed coarse mesh as a structural prior~\cite{habermann2021real,zhang_dynamic_2021,xiang2021modeling,zheng2024physavatar,li2024animatable,prokudin2023dynamic}.
Dynamic details are then attached to these templates through learned deformation networks and rendered to realistic appearance using neural rendering modules~\cite{habermann2021real,zhang_dynamic_2021,li2024animatable}.
Building on this idea, Gaussian-based representations~\cite{linLayGALayeredGaussian2024,li2024animatable} provide a more direct formulation that unifies 3D representation and rendering.
Beyond template-based pipelines, implicit representations have also been explored to model animatable avatars, by conditioning on dynamic parameters such as pose~\cite{tiwari2021neural,lee2023dynamic,wang2024pgahum} or skinning weights~\cite{saito2021scanimate}.
By integrating human priors as geometric constraints, methods like PGAHuman~\cite{wang2024pgahum} improve reconstruction fidelity by excluding empty regions from the occupancy field.
Together, these works demonstrate that pose-conditioned modeling is effective for capturing clothing dynamics that conform to body motion.

Another line of research focuses on geometry rather than appearance. Some works model avatars as point clouds, which is conceptually closer to our approach.
For instance, CloSET~\cite{zhang2023closet} decodes learned features bound to SMPL vertices into point clouds, while SCALE~\cite{ma2021scale} clusters points into local patches. Both enable animatable geometries with flexible topology.
However, their fidelity remains constrained by sparse sampling density, particularly in loose or high-frequency regions.
To address this, DPF~\cite{prokudin2023dynamic} learns a transition field from a canonical space to posed human space, allowing dense and continuous sampling. This concept inspires our approach, where we construct distribution between SMPL models and avatars.
In summary, while current approaches achieve high-fidelity avatars in unseen poses, they require video or 3D scan for each avatar, generally do not adopt a generative representation that supports flexible sampling across diverse avatars and motions.

\subsection{3D Avatar Generation}
Unlike animation modeling, generative methods require a compact representation for distribution learning, which poses challenges for high-fidelity synthesis.
A popular line of work employs GANs to synthesize tri-planes or NeRF-based representations for rendering~\cite{noguchiUnsupervisedLearningEfficient2022,zhangAvatarGen3DGenerative2022,yang2024attrihuman,wu20233dportraitgan,wu2024portrait3d}.
While effective for generating photorealistic appearance, these methods are constrained by rendering speed and resolution, often relying on auxiliary components such as super-resolution modules~\cite{bergman2022generative,dongAG3DLearningGenerate2023}, compositional structures~\cite{xuXAGen3DExpressive}, or refinement stages~\cite{menEn3DEnhancedGenerative2024,zhengSemanticHumanHDHighresolutionSemantic2024} to achieve high-fidelity results.
Incorporating human templates can alleviate sparse occupancy~\cite{hongEVA3DCompositional3D2022}, but challenges remain for loose clothing, and additional refinements are still commonly needed~\cite{huStructLDMStructuredLatent2024}.

Beyond radiance-field representations, several explicit formulations have been explored for static avatar generation.
Primitive volumes~\cite{chenPrimDiffusionVolumetricPrimitives2023} offer efficient rendering but face difficulties in synthesizing fine geometry.
Mesh-based methods, which predict displacement fields~\cite{sanyalSCULPTShapeconditionedUnpaired2023} or layered surface volumes~\cite{xu2024efficient}, often fail to capture complex topological variations such as garment wrinkles and loose dynamics. 
Point-based representations~\cite{abdalGaussianShellMaps2023,zhang$E^3$genEfficientExpressive2024,zhangHQavatarHighquality3D2024} provide flexibility for topology yet remain limited by point density and indirect supervision from 2D images. Based on Gaussian Splatting,  LHM~\cite{qiu2025lhm} learns animatable avatars from large-scale datasets, achieving notable generalization and diversity.
Building on this trend toward explicit geometry representation, geometry distribution~\cite{tang2025generative,zhang2024geometrydistribution} reformulates 3D as a distribution, enabling high-fidelity geometry synthesis from 3D data.
Despite these advances, most existing 3D avatar generation methods do not capture pose-dependent clothing deformations, resulting in static surface details that do not conform to body motion.

\section{Preliminaries}
\paragraph{Flow Matching.}
Flow matching~\cite{lipman2023flow} is a variant of diffusion models. It constructs a velocity field $u$ that transforms samples from a source distribution $\mathbf{p}$ to a target distribution $\mathbf{q}$. Specifically, the training objective loss is defined as follows.
\begin{equation}
\label{eq:flowmatching}
\arg\min_\theta\mathbb{E}_{\mathbf{x}_0\sim \mathbf{p}, \mathbf{x}_1\sim \mathbf{q}, t\in[0, 1]}\left\|u_\theta(\mathbf{x}_t\mid t) - (\mathbf{x}_1 - \mathbf{x}_0)\right\| 
\end{equation}
where $\mathbf{x}_t = (1-t)\mathbf{x}_0+t\mathbf{x}_1$. After training, an ordinary differential equation is solved to transition from $\mathbf{x}_0$ to $\mathbf{x}_1$, allowing us to sample $\mathbf{x}_1\sim \mathbf{q}$ by first sampling $\mathbf{x}_0\sim \mathbf{p}$. 

\paragraph{Human Geometry Distributions.}
\textit{Geometry Distributions}~\cite{zhang2024geometrydistribution} model a surface $\mathcal{M} \subset \mathbb{R}^3$ as a probability distribution $\Phi_\mathcal{M}$ (to avoid notation clutter, we also use the symbol $\mathcal{M}$ to denote the distribution), where any sample $\mathbf{x}_\mathcal{M} \sim \mathcal{M}$ corresponds to a point on the surface. Building on this idea, \textit{Human Geometry Distribution (HuGeoDis)}~\cite{tang2025generative} incorporates an additional SMPL surface $\mathcal{S}$. 
A naive idea is to use diffusion models to transform $\mathbf{x}_\mathcal{S}\sim\mathcal{S}$ to $\mathbf{x}_\mathcal{M}\sim\mathcal{M}$. However, HuGeoDis adopts a more efficient formulation by constructing a paired set $\mathbf{p} = \{(\mathbf{x}_\mathcal{S}, \mathbf{x}_\mathcal{M})\}$ and modeling the transformation from $\mathcal{N}(0, 1)$ to $\mathcal{T}(\mathbf{p})=\{\mathbf{x}_\mathcal{M} - \mathbf{x}_\mathcal{S}\mid (x_\mS,x_\mM) \sim \mathbf{p} \}$.
$\mathcal{T}$ works as the target distribution. The flow matching network is denoted by $u_\theta(\mathbf{x}_t\mid t,\mathbf{x}_\mS)$. It allows us to generate $\mathbf{x}_\mathcal{M}$ by conditioned on $\mathbf{x}_\mathcal{S}$.

\section{Method} 
For a single human avatar data sample, it is a SMPL-human pair $(\mathcal{S}, \mathcal{M})$. A dynamic sequence contains $N$ frames of pairs,
\begin{equation}
    \mathcal{H} = \{
    \overbrace{
        ((
        \underbrace{
            \mathcal{S}^{1}
        }_{\text{SMPL}}, 
        \underbrace{
            \mathcal{M}^{1}
        }_{\text{human}}
        ), (
        \underbrace{
            \mathcal{S}^{2}, \mathcal{M}^{2}
        }_{\text{a frame}}), \dots, (\mathcal{S}^{N}, \mathcal{M}^{N}) )
    }^{\text{a sequence}}
    \}.
\end{equation}
A dataset is composed of multiple sequences $\mathcal{D} = \{\mathcal{H}_1, \mathcal{H}_2, \dots, \mathcal{H}_M\}$ (we denote $\bar{\mathcal{D}}$ as the set of all frames). Our goal is to train a generative model of human sequences $\{\mathcal{M}^{1}, \mathcal{M}^{2}, \cdots, \mathcal{M}^{N}\}$ while conditioning on SMPL sequences $\{\mathcal{S}^{1}, \mathcal{S}^{2}, \cdots, \mathcal{S}^{N}\}$.

The training is a two-stage framework commonly used in latent diffusion models~\cite{rombach2021highresolution}. In the first stage, we encode all the frame data $\mathcal{M}$ into the latent space $\mathbf{z}$. We explain the basic encoding method~\cite{tang2025generative} in Sec.~\ref{sec:latent_space} and improve it in Sec.~\ref{sec:transition_construct} by introducing a new construction method of $\mathcal{T}$. In the second stage, we train generative models in the latent space $\mathbf{z}$. Specifically, it is a frame-by-frame autoregressive generative model while each frame is generated with flow matching (Sec~\ref{sec:animation_model}). Finally, we discuss the necessary data augmentation in Sec.~\ref{sec:data_aug}.

\begin{figure*}
    \centering
    \vspace{-2em}
    \includegraphics[width=1.0\linewidth]{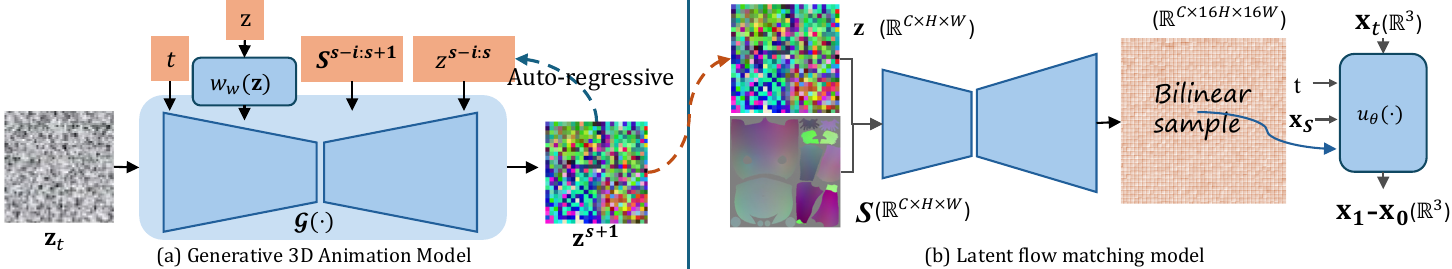}
    \captionsetup{skip=1pt} %
    \caption{(a) The animation model generates latent auto-regressively. (b) The latent flow-matching model samples detailed geometry from the latent space. }
     \label{fig:animation_model}
     \label{fig:latent_space}
     \vspace{-1em}
\end{figure*}

\subsection{Latent Space Modeling}
\label{sec:latent_space}
Following Tang \etal~\cite{tang2025generative}, we construct a latent space using HuGeoDis by compressing each geometry into a latent $\mathbf{z} \in \mathbb{R}^{C \times H \times W}$, represented by a rank-3 tensor with spatial dimensions $(H, W)$ and $C$ channels.  The latent, together with other geometry-specific cues, is incorporated as a condition of the flow matching model to assist in modeling diverse geometries. Note that the flow matching model here learns the geometry distributions, not the distribution across different identities. Specifically, the velocity model is represented as $u_\theta(\mathbf{x}_t\mid t, \mathbf{x}_\mS,\mathcal{S},\mathbf{z})$, where $\mathbf{x}_\mS\in \mathbb{R}^3$ denotes the corresponding point on the SMPL mesh, $\mathcal{S}$ is the SMPL mesh, encoding human pose and body shape, and $\mathbf{z}$ captures all remaining geometric information, such as avatar identity and clothing details.

Given a dataset $\bar{\mathcal{D}}$ where each avatar geometry is paired with an SMPL mesh, we assign a learnable latent $\textbf{z}$ for each pair, initialized with random Gaussian noise. This results in $\mathcal{Z} =\{(\mathcal{S}, \mathcal{M}, \mathbf{z}) | (\mathcal{S}, \mathcal{M})\sim\bar{\mathcal{D}} \}$.
We employ a small L2-regularization to the latent for training. Accordingly, the training objective can be formulated as:
\begin{equation}
\label{eq:hugeodis}
\begin{aligned}
     \min_{\theta,\{\mathbf{z}\}}
      \mathbb{E}_{\mathcal{Z}}
     & \mathbb{E}_{
            \mathbf{x}_0\sim\mathcal{N}, (\mathbf{x}_\mathcal{S},\mathbf{x}_\mathcal{M})\sim\mathbf{p}, t\in[0, 1]}
            \\
     & (
        \underbrace{
            \left\|u_\theta(\mathbf{x}_t\mid t, \mathbf{x}_\mS,\mathcal{S},\mathbf{z})-(\mathbf{x}_1-\mathbf{x}_0)\right\|
        }_{\text{flow matching loss}}
        +
        \underbrace{
            \beta\left\|\mathbf{z}\right\|_2
        }_{\text{regularizer}}
        ),
 \end{aligned}
\end{equation}
where $\mathbf{x}_1 = \mathbf{x}_\mM - \mathbf{x}_\mS.$
We follow the network architecture of Tang \etal~\cite{tang2025generative}. As illustrated in Fig.~\ref{fig:latent_space} (b), we represent $\mathcal{S}\in \mathbb{R}^{C\times H\times W}$ as a rank-3 tensor that shares the same resolution with the latent $\mathbf{z}$. A U-Net–based network takes $\mathcal{S}$ and $\mathbf{z}$ as inputs and outputs an upsampled feature map of size $\mathbb{R}^{C\times16H\times16W}$. $u_\theta$ then extracts features from this map by bilinear sampling to integrate information from both $\mathcal{S}$ and $\mathbf{z}$. As discussed above, $u_\theta$ allows us to predict $\mathcal{M}$ by conditioning on $\mathcal{S}$.

\subsection{Low-cost Mapping Construction}
\label{sec:transition_construct}

\begin{figure}
    \centering
    \includegraphics[width=0.9\linewidth]{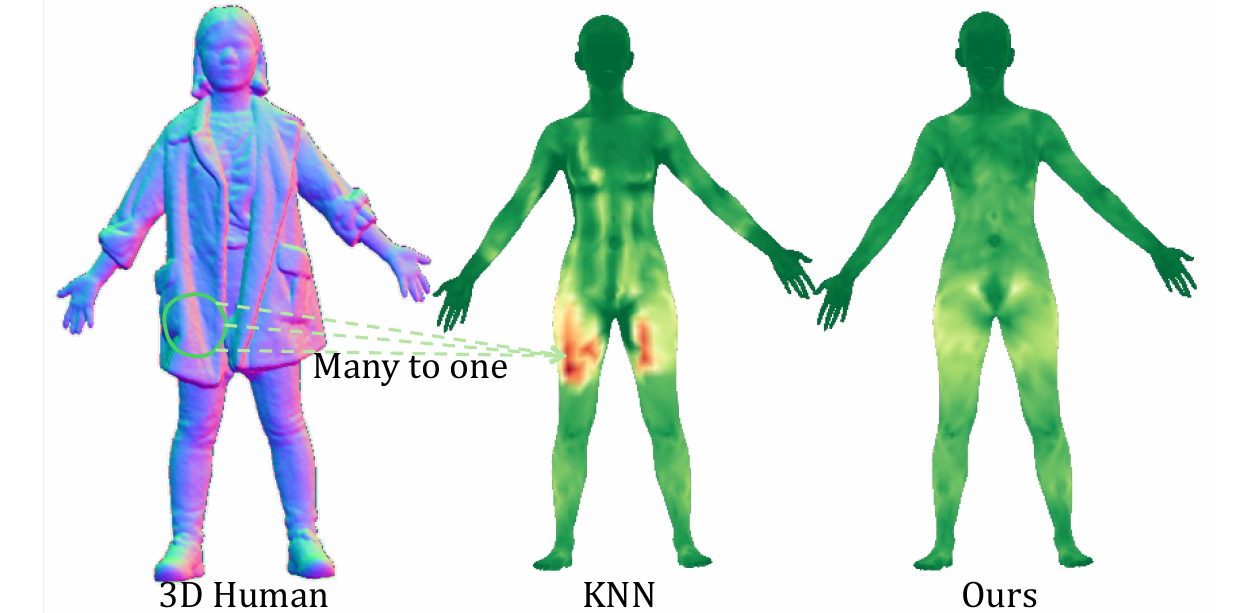}
    \caption{Visualization of the density distribution of mapped points. The green indicates fewer points, red indicates more.}
    \label{fig:transition_density}
    \vspace{-1em}
\end{figure}

We model the transformation from $\mathcal{N}(0, 1)$ to $\mathcal{T}(\mathbf{p})=\{\mathbf{x}_\mathcal{M} - \mathbf{x}_\mathcal{S}\mid (\mathbf{x}_\mS,\mathbf{x}_\mM) \sim \mathbf{p} \}$, which requires sampling ($\mathbf{x}_\mS,\mathbf{x}_\mM)$ from a paired set $\mathbf{p}$, as shown in Eq.~\ref{eq:hugeodis}. Here, $\mathbf{p}$ defines a probabilistic mapping between $\mS$ and $\mM$. While HuGeoDis~\cite{tang2025generative} demonstrates that constructing $\mathbf{p}$ with low-cost pairs (e.g., $\mathbf{x}_\mM$ spatially close to $\mathbf{x}_\mS$) significantly improves training efficiency, it does not provide a mapping that is sufficiently low-cost and reliable. Specifically,
HuGeoDis employs the KNN algorithm to find the nearest point on the SMPL mesh for each point on the 3D human, denoted by:
\begin{equation}
  \{(\mathbf{x}_\mM,\mathbf{x}_\mS)\mid \mathbf{x}_\mM\sim \mM,\mathbf{x}_\mS=\arg\min_{\mathbf{x}_\mS'\sim\mS}||\mathbf{x}_\mM - \mathbf{x}_\mS'||\}
\end{equation}

However, this produces an uneven distribution of points on the SMPL mesh: some $\mathbf{x}_\mS$ points are associated with a large number of $\mathbf{x}_\mM$, while others have few or no correspondences, as shown in Fig~\ref{fig:transition_density}.
As a result, training on such imbalanced mapping densities leads to suboptimal geometry reconstruction, since uniformly sampling from $\mS$ yields uneven sampling on $\mM$, resulting in incomplete or hole-prone geometries unless a large number of samples are drawn. See Fig.~\ref{fig:recon_compare} (HuGeoDis) for a visualization.

The optimal way to construct the mapping would be through optimal transport between the 3D human and the SMPL mesh. However, this approach is computationally expensive and would need to be computed for each 3D human in the training dataset.
Instead, we propose an alternative solution that relies on the computation of an efficient and approximate mapping between the SMPL mesh $\mathcal{S}$ and the 3D human $\mathcal{M}$. We compute the approximate mapping $m_\phi(\mathbf{x}_\mS\mid \mS, \mathbf{z}_m)$ by training a supervised model to obtain coarse yet uniform correspondences between the SMPL and 3D human surfaces. This mapping is deterministic and one-to-one. 
We adopt a similar network structure as $u_\theta(\mathbf{x}_t\mid t, \mathbf{x}_\mS,\mathcal{S},\mathbf{z})$, denoted as $m_\phi(\mathbf{x}_\mS\mid \mS, \mathbf{z}_m)$. Here, $\mathbf{z}_m$ is also initialized as random noise and optimized during training optimizing:
\begin{equation}
\label{eq:transitions}
\begin{aligned}
     \min_{\phi}
      \mathbb{E}_{\mathcal{Z}_m}\mathbb{E}_{\mathbf{x}_\mathcal{S}\sim\mathcal{S}, \mathbf{x}_\mathcal{M}\sim\mathcal{M}}\text{Chamfer}(m_\phi(\mathbf{x}_\mS\mid \mS,\mathbf{z}_m),\mathbf{x}_\mM),
 \end{aligned}
\end{equation}
where $\mathcal{Z}_m=\{\mS,\mM,\mathbf{z}_m\}$ and the loss is Chamfer distance.
After training $m_\phi$, we construct $\mathbf{p}$ based on the deterministic map $m_\phi$ (see appendix for details), and then train $u_\theta$ using the flow matching model to learn the distribution $\mathcal{T}(\mathbf{p})$.

Fig.~\ref{fig:transition_density} visualizes the mapping.
Regions such as the face and hands exhibit nearly one-to-one correspondences due to minimal geometric differences between the SMPL and 3D human, whereas loose clothing areas show a higher density of mapped points. Our mapped points are distributed more uniformly than those obtained by directly applying KNN.


\subsection{Generative 3D Animation Model}
\label{sec:animation_model}
The encoding stage learns a latent for each data sample, thus our dataset becomes
\begin{equation}
    \{((\mathcal{S}^{1},\mathcal{M}^{1}, \mathbf{z}^{1}), (\mathcal{S}^{2}, \mathcal{M}^{2}, \mathbf{z}^{2}),
        \dots, (\mathcal{S}^{N}, \mathcal{M}^{N}, \mathbf{z}^{N}))\}.
\end{equation}
We train a conditional diffusion model to generate human animations in the latent space $\mathbf{z}$ based on the analysis of three design choices. 1) Due to the scarcity of 3D human animation data, directly modeling long sequences is challenging, as it requires learning complex temporal dependencies from limited samples (See Sec.~\ref{sec:animation_model}). Therefore, we decompose the animation generation into a sequence of short-duration transitions and formulate it as an autoregressive process. By maintaining a fixed short-term context of $i$ time-steps, the model can efficiently generate sequences of arbitrary length without increasing computational complexity. 2) An auto-regressive model by itself introduces drift in the human geometry. We need to add additional conditioning information encoding the human geometry of a single fixed time step. 3) Data augmentation is needed to avoid overfitting. Otherwise, the model learns to associate a specific pose with a single human geometry.

Specifically, the short-term dynamics are modeled by a flow matching model $v_\psi(\mathbf{z}_t | t, \mathbf{z}^{s-i: s}, \mathcal{S}^{s-i: s+1},c)$, which can infer the next-frame latent $\mathbf{z}^{s+1}$ from the past latents $\mathbf{z}^{s-i: s}=\{\mathbf{z}^{s-i},...,\mathbf{z}^s\}$ and the corresponding poses $\mathcal{S}^{s-i: s+1}=\{\mS^{s-i},...,\mS^{s+1}\}$. 
To support long-term consistency, we introduce an additional condition $c$ encoding avatar identity and clothing, preserving appearance across frames.  The training objective loss $\mathcal{L}_{\text{diff}}$ of diffusion model is:
\begin{equation}
\label{eq:hugeodis}
\begin{aligned}
     \min_{\psi} \mathbb{E}_{\mathcal{D}} & \mathbb{E}_{s\in[1, N], \mathbf{n}\sim\mathcal{N}, t\in[0, 1]} \\
      & \left\|v_\psi(\mathbf{z}_t|t, \mathbf{z}^{s-i: s}, \mathcal{S}^{s-i: s+1},c)-(\mathbf{z}^{s+1}-\mathbf{n})\right\|,
 \end{aligned}
\end{equation}
where $\mathbf{z}_t = (1-t)\mathbf{n} + t\mathbf{z}^{s+1}$.

We use a downsampled convolutional model $w_\omega(\mathbf{z})$ to obtain the condition $c$. During training, $c$ is computed from a latent at an arbitrary frame of the same sequence. During inference, we compute $c$ from the first frame $\mathbf{z}^0$ and keep it fixed throughout the entire sequence. The network $w_\omega(\mathbf{z})$ is trained with contrastive learning using the NT-Xent loss~\cite{sohn2016improved}, which encourages $c$ to distinguish between different avatars and clothing styles. The loss $\mathcal{L}_{\text{nt-xent}}$ is defined as follows:
\begin{equation}
    \min_\omega-\frac{1}{N}\sum_{(i,j) \in \mathcal{A}}\text{log}\frac{\text{exp}(\text{sim}(c^i,c^j)/\tau)}{\sum_{k\neq i}\text{exp}(\text{sim}(c^i,c^k)/\tau)},
\end{equation}
where $c^i = w_\omega(\mathbf{z}^i)$, $\mathcal{A}$ is the set of positive pairs (frames of the same avatar and clothing), $N$ is the number of positive pairs, $\text{sim}(\cdot,\cdot)$ denotes cosine similarity, and $\tau$ is a temperature factor. Therefore, the loss of our generative animation model can be represented by:
\begin{equation}
    \mathcal{L} = \mathcal{L}_{\text{diff}} + \alpha\mathcal{L}_{\text{nt-xent}},
\end{equation}
where $\alpha$ is a weight factor. 
As illustrated in Fig.~\ref{fig:animation_model} (a), we adopt a U-Net architecture for the diffusion model $v_\psi$. During inference, the predicted latent $\mathbf{z}^{s+1}$ is appended to the past latents $\mathbf{z}^{s-i:s}$ to synthesize the animation sequence progressively.  Inspired by classifier-free guidance, we randomly replace conditions with null embeddings during training, enabling the synthesis of the first few frames where past latents are not available.

Note that the condition $c$ plays a crucial role in our framework by providing information beyond short-term durations, 
thereby preventing the forgetting problem commonly observed in long-sequence video models. 
Comparisons with the model without condition $c$ are presented in Sec.~\ref{sec:exp_animation}.

\subsection{Data Augmentation}
\label{sec:data_aug}
To enable our model to generalize to different body shapes and expressions, we apply data augmentation on $\mathcal{S}$ by varying its body shape and expression parameters during training of $u_\theta$ and $v_\psi$. To ensure the augmented parameters remain reasonable, we interpolate them with the standard no-expression body template using a random factor in the range $(-1.0, 1.5)$. Here, $0$ corresponds to the standard template, and $1$ corresponds to the original parameters. Negative values produce inverse deformations (e.g., transforming a fat shape to thin), while positive values greater than $1$ produce exaggerated deformations. After interpolation, we randomly shuffle these parameters within each batch. 

Data augmentation helps disentangle the appearance latent $\mathbf{z}$ from the SMPL mesh $\mathcal{S}$. Without this, $u_\theta$ may memorize correlations between $\mathcal{S}$ and appearance, leading to inconsistent synthesis when the SMPL parameters are varied.
\section{Experiments}
We conduct experiments to validate our proposed 3D latent space and animation generation framework. We evaluate three aspects: (i) reconstruction accuracy and efficiency of $u_\theta(\cdot)$ (Sec.~\ref{sec:exp_recon}), (ii) generation performance on the latent space, following Tang \etal~\cite{tang2025generative} (Sec.~\ref{sec:exp_3dgeneration}), and (iii) quality of the generated animations (Sec.~\ref{sec:exp_animation}).
\subsection{Reconstruction}
\label{sec:exp_recon}
Our proposed low-cost mapping enables more uniform and efficient sampling, allowing the model to reconstruct high-fidelity geometry using fewer sampled points.

\begin{figure}[t]
    \centering
     \vspace{-1em}
    \includegraphics[width=1.0\linewidth]{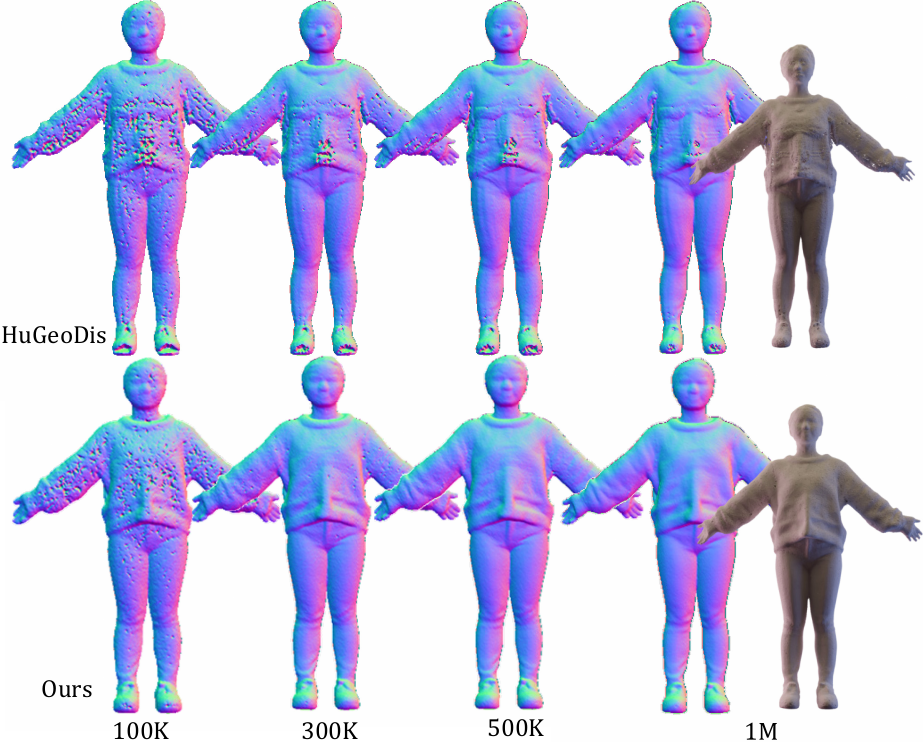}
    \caption{Comparison across different number of sampling. We apply GS-rendered normal maps for superior detail visualization, which may slightly inflate boundaries due to non-zero GS scales; a point cloud rendering is shown (right) for boundary reference.}
    \label{fig:recon_compare}
    \vspace{-1em}
\end{figure}

To validate the reconstruction quality, we measure the Chamfer Distance between the reconstructed 3D avatars and the ground truth on the 4d-dress~\cite{wang20244ddress} dataset. We also report the time cost for sampling a different number of points with 20 denoising steps on an NVIDIA A100 GPU. Both results are summarized in Tab.~\ref{tab:recon_acc}.
Across different sampling settings, our method achieves substantially higher reconstruction accuracy, reducing the Chamfer Distance by up to an order of magnitude compared to the original HuGeoDis~\cite{tang2025generative}.
The qualitative comparisons in Fig.~\ref{fig:recon_compare} further confirm this advantage: our approach produces smoother and higher-fedility surfaces with far fewer sampling points. Notably, it can faithfully capture the full geometry using around 300K points, while HuGeoDis fails to achieve full coverage even with 1 million points. As a result, the reduced number of sampling points leads to a substantial improvement in computational efficiency. 

For reference, we also evaluate the supervised model $m_\phi$ used for computing $\mathcal{T}$, which provides coarse yet uniform correspondences between the SMPL mesh and 3D human surfaces, as described in Sec.~\ref{sec:transition_construct}. Although it shares a similar architecture with our flow matching network, it yields significantly lower reconstruction accuracy, underscoring the importance of a distribution-based representation.

\begin{table}[h]
\centering
\captionsetup{font=small}
\caption{Comparison of Chamfer distance ($\times 10^{-5}$) and time cost (seconds) under different sampling points.}
\small
\setlength{\tabcolsep}{6pt} 
\begin{tabular}{lcccc}
\toprule
Method & 100K & 300K & 500K & 1M \\
\midrule
HuGeoDis~\cite{tang2025generative}    & 2.65 & 2.09 &1.95 & 1.86 \\
Supervised    & 2.72 & 2.48 & 2.43 & 2.42\\
Ours    & \textbf{0.52}  & \textbf{0.27} & \textbf{0.22} &\textbf{0.15} \\ 
\midrule
Time (s) & 2.15  & 6.22 & 10.28 & 20.59  \\
\bottomrule
\end{tabular}
\vspace{-1em}
\label{tab:recon_acc}
\end{table}

\subsection{Static Random 3D Human Generation}
\label{sec:exp_3dgeneration}
Generating diverse and high-quality 3D human geometries is a fundamental task in avatar generation. To assess the generative capability of our latent space, we follow Tang \etal~\cite{tang2025generative} and train a random generation model using our proposed latent space. The generation quality is measured by the Fréchet Inception Distance (FID) computed between the rendered normal maps of generated results and the ground-truth normals. We use the THuman2 dataset~\cite{tao2021function4d} for both training and evaluation. For each subject, we render $50$ views and collect a total of $25,000$ normal images across the dataset.

We compare our approach with representative human generation methods that adopt different 3D representations, including Gaussian splatting (E3Gen~\cite{zhang$E^3$genEfficientExpressive2024}), implicit functions (GetAvatar~\cite{fengLearningDisentangledAvatars2023}, gDNA~\cite{chenGDNAGenerativeDetailed2022}), neural radiance fields (ENARF~\cite{noguchiUnsupervisedLearningEfficient2022}, GNARF~\cite{bergman2022generative}, Eva3D~\cite{hongEVA3DCompositional3D2022}), and geometry-distribution-based representations (HuGeoDis~\cite{tang2025generative}).
Some of these methods~\cite{chenGDNAGenerativeDetailed2022,zhangGETAvatarGenerativeTextured2023,zhang$E^3$genEfficientExpressive2024} employ auxiliary cues such as normal maps or predicted point rotations to refine the rendering normal quality. Since our focus is on geometry fidelity, we primarily compare their raw geometry outputs, while also reporting FID scores of their enhanced rendered results for completeness.

As shown in Tab.~\ref{tab:FID_normal}, geometry-distribution-based representations (HuGeoDis and ours) consistently achieve superior results in raw geometry comparisons, significantly improving over alternative 3D representations. Furthermore, our method surpasses HuGeoDis~\cite{tang2025generative} with higher geometric quality due to our better mapping construction.
\begin{table}[h]
\centering
\vspace{-0.5em}
\captionsetup{font=small}
\caption{Comparison of FID scores. 
The * results are adopted from E3Gen~\cite{zhang$E^3$genEfficientExpressive2024}. 
For some methods, the raw and enhanced renderings are identical.}
\small
\setlength{\tabcolsep}{6pt} 
\begin{tabular}{lcc}
\toprule
Method & Raw Geometry & Enhanced Rendering \\
\midrule
ENARF*~\cite{noguchiUnsupervisedLearningEfficient2022}    & 223.72 & 223.72 \\
GNARF*~\cite{bergman2022generative}    & 166.62 & 166.62 \\
EVA3D*~\cite{hongEVA3DCompositional3D2022}    & 60.37  & 60.37  \\
E3Gen~\cite{zhang$E^3$genEfficientExpressive2024}     & 65.32  & 28.12  \\
GetAvatar~\cite{fengLearningDisentangledAvatars2023} & 56.07  & 22.77  \\
gDNA~\cite{chenGDNAGenerativeDetailed2022}      & 42.90  & 17.43  \\
HuGeoDis~\cite{tang2025generative}  & 16.16  & 16.16  \\
Ours      & \textbf{14.03} & \textbf{14.03} \\
\bottomrule
\end{tabular}
\vspace{-1.5em}
\label{tab:FID_normal}
\end{table}

\subsection{Animation Generation}
\label{sec:exp_animation}
Compared to existing avatar generation methods~\cite{qiu2025lhm,zhang$E^3$genEfficientExpressive2024}, the key advantage of our approach lies in its ability to generate natural clothing deformations with fine-grained geometric details.
We evaluate our models on 4d-dress~\cite{wang20244ddress} dataset in three aspects: geometric quality, identity consistency and deformation naturalness.
Geometric quality is evaluated by the FID metric computed on rendered normal images, same as Sec.~\ref{sec:exp_3dgeneration}.
For identity consistency evaluation, we train an ID classifier based on the DINO network~\cite{oquab2023dinov2} (details provided in the Appendix). For each generated sequence, we compute the cosine similarity of identity features between every frame and the first frame to measure temporal consistency.
Deformation naturalness is evaluated through a user study, as no standard quantitative metric exists.
Each participant provides three scores: one assessing the quality of the geometry, another evaluating the naturalness of clothing deformation, without considering the human motion, and a third measuring how well the deformation conforms to body motion.
We evaluate each method on 90 generated sequences, covering 6 identities and 13 motions per identity, along with several randomly generated samples, resulting in approximately 6,000 frames for each method. 
Both the comparison and ablation results are presented in Tab.~\ref{tab:animation_comp}.

\paragraph{Comparisons.}
\vspace{-1em}
We compare our method with LHM~\cite{qiu2025lhm} and a long-term supervision model that follows the auto-regressive framework~\cite{tang2022real,tang2023rsmt}, which explicitly learn long range dependencies (see Appendix for details). The LHM learns Gaussian splatting (GS) animatable avatars from large-scale datasets and takes an image as input. To focus on geometry and ensure a fair comparison, we evaluate their results in two settings. In the first setting, we use the normal map of an avatar as the input image, treating it as a color image. This allows LHM to preserve as much of the input normal details as possible but does not reveal the underlying geometry encoded in the GS representation. In the second setting, we ignore color information and extract the real normals from the GS depth map, which provides a faithful representation of the actual geometry. 

\begin{table}[h]
\centering
\vspace{-0.5em}
\captionsetup{font=small}
\caption{Comparison and ablation study of generated animations.}
\small
\resizebox{\linewidth}{!}{
\begin{tabular}{lccccc}
\toprule
Method & FID $\downarrow$ & ID $\uparrow$ &  Quality $\uparrow$ & Naturalness  $\uparrow$ & Comformance  $\uparrow$ \\
\midrule
LHM (normal)    & 33.37   & N/A & 3.3 &2.7 & 2.0 \\
LHM (geometry)    & 58.19   &   N/A &1.3 &1.7 & 1.5\\
long-term    & 27.13   &  0.61 & 3.0&2.2 & 2.5\\
\midrule
w/o condition    &  27.68  & 0.60 &3.1&2.2 & 2.6 \\
w/o augment  &\textbf{24.20}  &0.76 &3.7 & 3.1 & 3.5 \\
\midrule
Ours      &\underline{25.01}&\textbf{0.96}  & \textbf{4.4} &\textbf{4.5}&\textbf{4.4}  \\
\bottomrule
\end{tabular}}
\label{tab:animation_comp}
\end{table}

\begin{figure*}
    \centering
    \vspace{-2em}
    \includegraphics[width=1.0\linewidth]{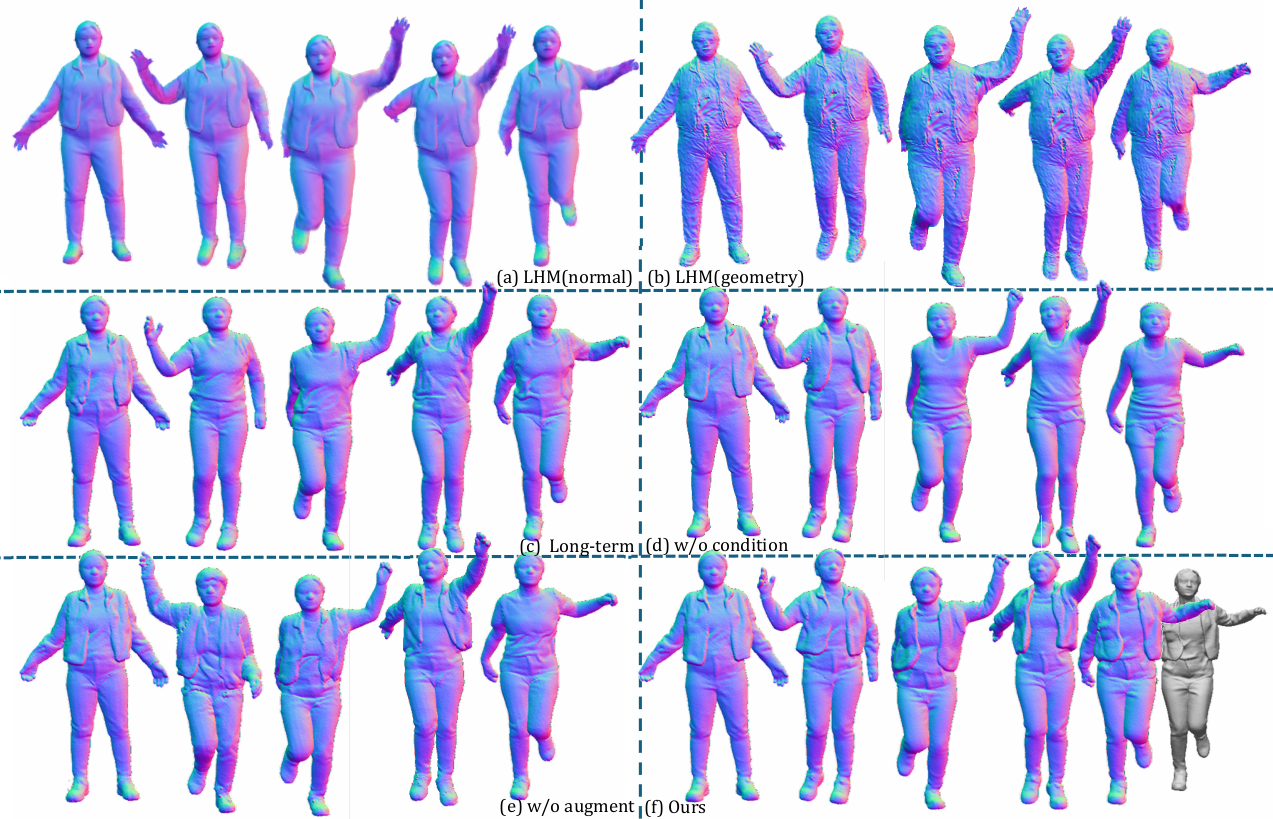}
    \caption{Results for a female (the identity corresponding to the first frame of each sequence) wearing an outer garment while running and swinging her arms. The garment deforms naturally and follows the body movements. Our approach is the only one that simultaneously captures the dynamic behavior of the garment, preserves high-quality geometric details, and consistently maintains the identity. A reconstructed mesh is shown in (f) for boundary reference.}
    \label{fig:animation_compare}
    \vspace{-1.5em}
\end{figure*}

As seen in Tab.~\ref{tab:animation_comp}, our method achieves competitive results across all metrics. Specifically, in terms of the FID metric, all distribution-based models (long-term, ours) outperform LHM significantly, indicating a stronger capacity to represent accurate geometry. Moreover, regarding the dynamic aspect, our method achieves the best ID score and the highest user study scores in terms of quality, naturalness, and conformance. We omit LHM results from quantitative ID evaluation because its generated normals fall outside the distribution of our ID classifier, making the evaluation unreliable. Besides, the synthesized appearance of LHM is deformed via rigging. While this approach preserves identity trivially, it fails to capture realistic garment dynamics, as evidenced by its lower user study scores and the visual results shown in Fig.~\ref{fig:animation_compare}~(a,b), where the clothing details remain unchanged. 
As for the “long-term” model, it tends to overfit to training motions, producing low-quality and inconsistent results for unseen movements (Fig.~\ref{fig:animation_compare} (c)).
\paragraph{Ablation study.}
\vspace{-1em}
We include two models for ablation study:
(1) a model without long-term consistency condition (w/o condition);  and 2) the model trained without augmented dataset (w/o augment).

As shown in Tab.~\ref{tab:animation_comp}, although both our full model and the “w/o augment” model maintain high-fidelity generation in terms of FID score, the latter exhibits appearance leakage correlated with specific poses. As a result, it achieves a slightly lower FID (closer to the dataset distribution) but suffers from reduced consistency across poses, as reflected by worse ID scores and identity inconsistencies in intermediate frames (Fig.~\ref{fig:animation_compare} (e)). Furthermore, the “w/o condition” model gradually drifts away from the original identity over time (Fig.~\ref{fig:animation_compare} (d)), highlighting the importance of conditioning on avatar identity for maintaining temporal stability. 
 Overall, these results demonstrate that our approach uniquely preserves high-fidelity geometric details while producing natural and temporally coherent dynamics. More qualitative comparisons are provided in the Appendix.

\section{Limitations and Future Work}

The scarcity of high-quality 3D human animation data poses a major challenge to our work. To learn and generalize garment dynamics across different clothing types, a dataset must contain not only sufficient static geometry but also diverse clothing motions. Despite the limited data available, our method still demonstrates strong generalization to different garments and identities. However, this data scarcity inevitably leads to several limitations:
(1) While synthesizing physically plausible results, our method does not guarantee physically accurate animations. Since unseen dynamics are generalized from garments of different materials, the generated motions may be inconsistent with the current clothing type (e.g., a stiff jacket exhibiting soft, fabric-like dynamics).
(2) Minor garment–body intersections may occasionally occur, particularly in unseen poses. In this regard, explicitly handling interpenetrations could further improve physical realism.
We expect that with larger and more diverse datasets, the physical realism of the generated animations will improve. Future work may also consider decomposing the avatar and garments, enabling fine-grained editing of the generated animations.
\section{Conclusion}
We propose a generative framework for 3D human geometry animation that jointly addressed two key challenges: preserving fine-grained geometric details and modeling natural clothing dynamics under limited data. Our two-stage framework first learns a latent space and then trains a generative model on it. Fine-grained clothing details are achieved through our human geometry distribution, which enables more uniform sampling than existing methods, leading to higher-quality and more efficient generation. To model temporal dependencies from limited animation data, we train a conditional diffusion model on short-term dynamics while incorporating long-term consistency through a conditional input. We conduct experiments to evaluate both our latent space and our animation generation framework. Our latent space provides superior reconstruction and generation performance. Our framework produces temporally coherent human geometry with detailed and natural clothing dynamics.
{
    \small
    \bibliographystyle{ieeenat_fullname}
    \bibliography{main}
}

\clearpage
\setcounter{page}{1}
\maketitlesupplementary

\section{Implementation Details}
\subsection{Low-cost Mapping Construction}
\begin{figure}
    \centering
    \includegraphics[width=1.0\linewidth]{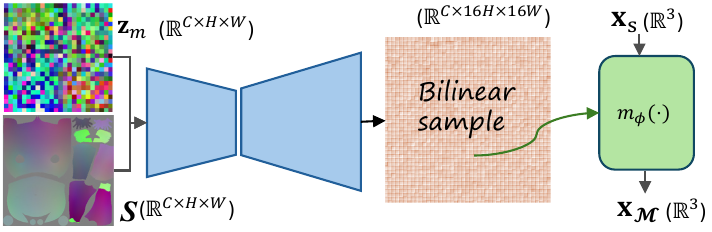}
    \caption{The illustration of the supervised model.}
    \label{fig:supervsied}
\end{figure}

We adopt a similar structure as $u_\theta(\mathbf{x}_t\mid t,\mathbf{x}_\mS,\mS,\mathbf{z})$ to train $m_\phi$. Instead of using a diffusion model, we train $m_\phi$ as a supervised model for efficiency. As shown in Fig.~\ref{fig:supervsied}, we replace the input of $u_\theta$ with $\mathbf{x}_\mS$, and output the point $\mathbf{x}_\mM$ on the target geometry directly. Accordingly, we discard the condition branch of $\mathbf{x}_\mS$ and $t$. 

After training $m_\phi$, we construct $\mathcal{T}$ via three steps.

\textbf{Step 1. Coarse mapping.} We first uniformly sample millions of points from both SMPL mesh $\mathbf{x}_\mS \sim \mS$ and 3D human geometry $\mathbf{x}_\mM \sim \mM$,  and use the trained model to obtain a coarse prediction $m_\phi(\mathbf{x}_\mS)$ for each $\mathbf{x}_\mS$.

\textbf{Step 2. Initial transitions.} For each $\mathbf{x}_\mM$, we find its nearest $m_\phi(\mathbf{x}_\mS)$ using KNN to determine its corresponding $\mathbf{x}_\mS$, yields the initial transitions:
\begin{equation}
    \mathcal{T}_0 = \{(\mathbf{x}_\mS,\mathbf{x}_\mM)\mid \mathbf{x}_\mS=\arg\min_{\mathbf{x}_\mS'}||\mathbf{x}_\mM - m_\phi(\mathbf{x}_\mS')||\}
\end{equation}
However, since not all $\mathbf{x}_\mS \in \mS$ are selected as nearest neighbors, some SMPL mesh regions remain sparsely covered or entirely unsampled. This incomplete coverage leads to under-trained regions on the SMPL mesh and inaccurate estimation during inference.

\textbf{Step 3. Coverage refinement.} 
To address this issue, we first identify triangle faces on the SMPL mesh that lack any sampled points from $\mathcal{T}_0$. 
Let $\mathcal{F}_{\rm covered}$ denote faces that contain at least one $\mathbf{x}_\mS$ in $\mathcal{T}_0$, and $\mathcal{F}_{\rm uncovered} = \mathcal{F} \setminus \mathcal{F}_{\rm covered}$ the remaining faces.
We uniformly sample new points on $\mathcal{F}_{\rm uncovered}$, estimate their target locations $m_\phi(\mathbf{x}_\mS)$, 
and find their nearest $\mathbf{x}_\mM$ on the human surface. 
These additional transitions, $\mathcal{T}_{\rm add}$, are merged with $\mathcal{T}_0$ to produce the final set 
$\mathcal{T} = \mathcal{T}_0 \cup \mathcal{T}_{\rm add}$, 
ensuring better coverage and a more uniform transitions.
\subsection{Training Configurations}
During training of the latent space, we adopt a network architecture similar to HuGeoDis~\cite{tang2025generative}. Specifically, we sample $2^{18}$ training transitions for each geometry, and the latent feature $\textbf{z}$ is represented as $\mathbb{R}^{6\times24\times24}$. The model employs both linear and convolutional layers from EDM2~\cite{Karras2024edm2}, with 256 channels used in all layers.

For the generative animation model $\mathcal{G}$, we employ a U-Net architecture following EDM2~\cite{Karras2024edm2}. The network performs three downsampling operations, with channel dimensions of $384$, $768$, and $1536$ for each respective stage. Attention layers are applied at every resolution level.
The conditioning input of animation model is provided by a convolutional network that takes the latent feature $\textbf{z}$ as input. Specifically, $\phi(\cdot)$ has a channel dimension of 32 and spatial size $24\times24$, which is progressively downsampled through three convolutional layers to a resolution of $3\times3$, yielding $\phi(\textbf{z})\in \mathbb{R}^{32\times 3\times 3}$.

For optimization, we use AdamW~\cite{loshchilov2017decoupled} with a learning rate of $8\times10^{-3}$, and apply a cosine learning rate scheduler across all models. A small L2 regularization of $10^{-6}$ is applied to the latent features. We observe that stronger regularization tends to make the network rely excessively on the SMPL condition, leading to potential overfitting to SMPL-driven patterns.

The latent model is trained for approximately five days using four NVIDIA A100 GPUs, while the generative animation model is trained for two days on the same hardware configuration. The batch size is set to 16 per GPU.
\subsection{Dino-based Identity Classifier}
To evaluate identity similarity, we train an identity classifier based on the DINO network.
Given a normal image as input, the DINO encoder converts it into a sequence of tokens, consisting of one [CLS] token and multiple image feature tokens.
These tokens are then processed by $6$ Transformer blocks, followed by an MLP head that maps the [CLS] token into a $64$-dimensional embedding.
The network is trained with a contrastive learning objective using the NT-Xent loss, which minimizes the distance between embeddings of the same identity and maximizes the distance between embeddings of different identities.
\subsection{Long-term Supervision Model}
We build a non-diffusion baseline following an auto-regressive framework commonly used in the motion domain~\cite{tang2023rsmt} for long-term supervision. Unlike the diffusion-based model, which learns denoising at each step, this supervised model learns to predict clean latents directly, enabling efficient long-range generation. Specifically, the model directly predicts the next latent state $\mathbf{z}^{s+1}$ conditioned on the known context $\{\mathbf{z}^{s-i:s}, \mathcal{S}^{s-i:s+1}, c\}$, rather than predicting noise or score.

During training, the model auto-regressively generates a sequence of $n=8$ animation steps, and supervision is applied over all generated frames:
\begin{equation}
    \mathcal{L}_{\text{sup}} = 
    \sum_{s=1}^{n=8} \left\| \hat{\mathbf{z}}^{s} - \mathbf{z}^{s} \right\|_2^2.
\end{equation}
This design allows direct long-horizon supervision and helps mitigate the accumulation of single-step prediction errors when sufficient training data is available.

\section{Additional Results}
\begin{figure*}[t]
    \centering
    \vspace{-2.5em}
    \includegraphics[width=1.0\linewidth]{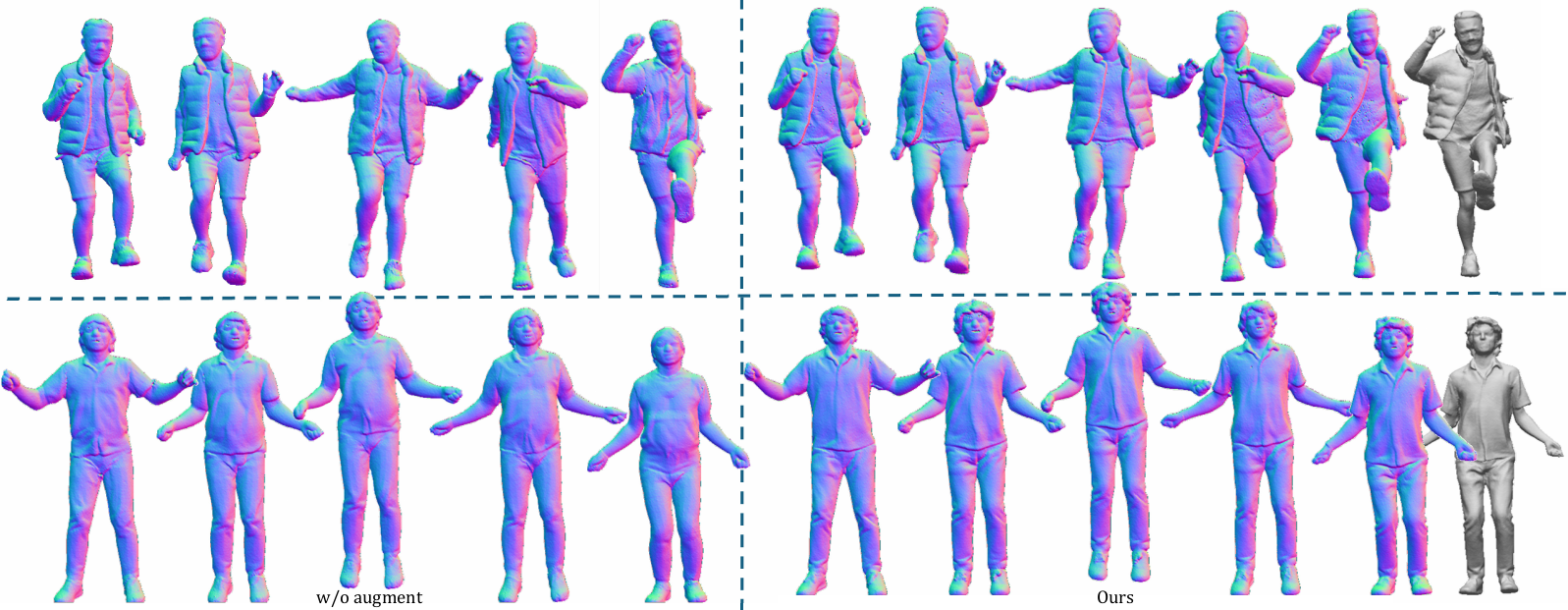}
    \captionsetup{skip=1pt} %
    \caption{Comparison between our model and the ``w/o augment'' model. The first rwo shows a leg-kicking motion, where the outer garment naturally follows the body movement, while the second row depicts a rope-jumping motion. The ``w/o augment'' model tends to generate garments associated with other character identities, indicating identity leakage across sequences.}
    \label{fig:no_aug_compare}
\end{figure*}

\begin{figure*}[h]
    \centering
      \vspace{-1em}
    \includegraphics[width=1.0\linewidth]{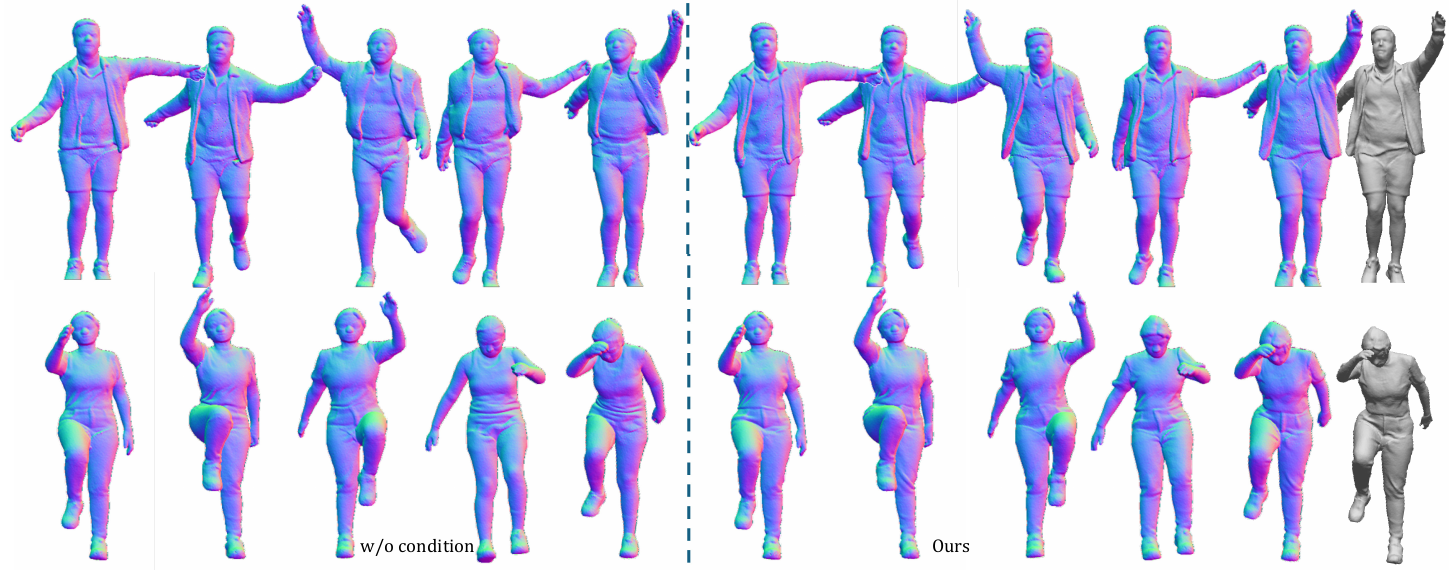}
        \captionsetup{skip=1pt} %
    \caption{Comparison between our model and the ``w/o condition'' model. The first row shows running with swinging arms, and the second row shows leg lifts with arm swings. Without long-term consistency, the generated character identity gradually changes over time.}
    \label{fig:id_compare}
\end{figure*}

\begin{figure*}[h]
    \centering
      \vspace{-1em}
    \includegraphics[width=1.0\linewidth]{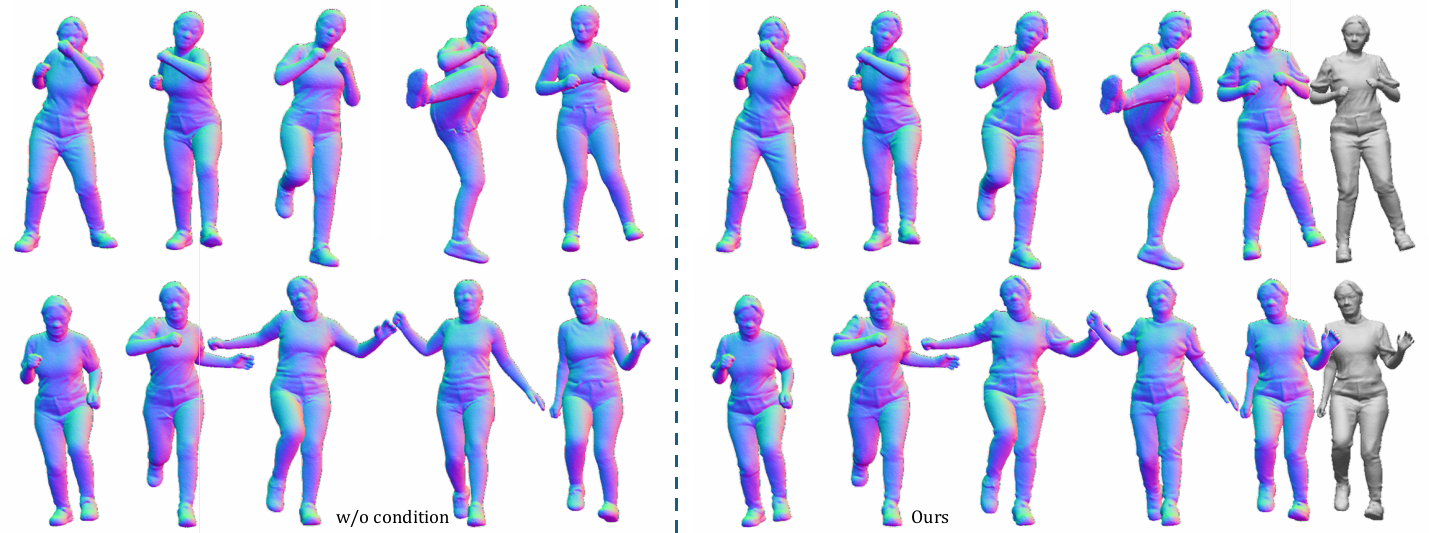}
    \caption{Comparison between our model and the ``w/o condition'' model. The first row shows martial arts kicking motions, and the second row shows running with arm swings. Without long-term consistency, the generated identity gradually drifts over time. }
    \label{fig:id_compare2}
\end{figure*}

\begin{figure*}[h]
    \centering
    \vspace{-5em}
    \includegraphics[width=0.95\linewidth]{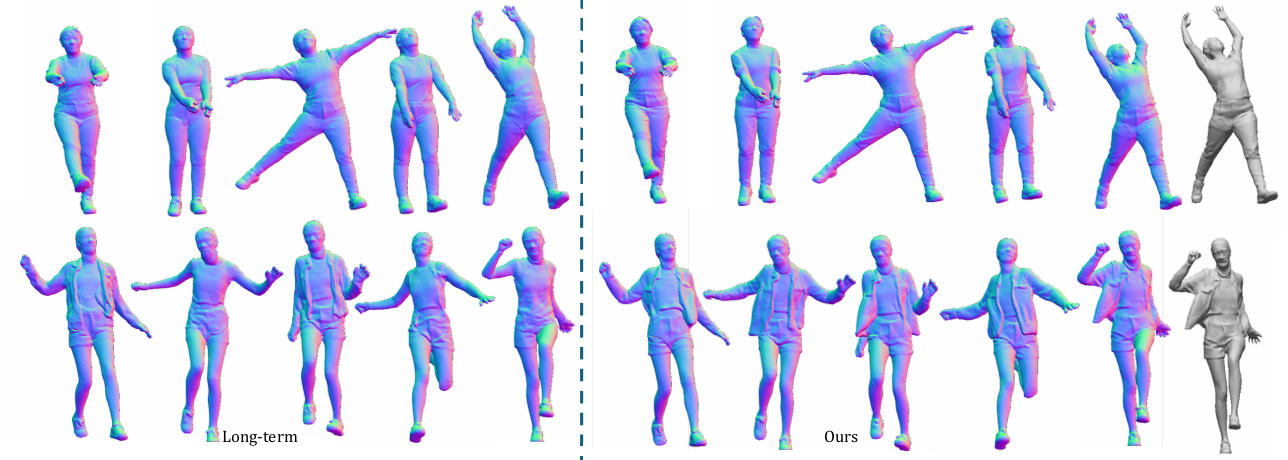}
    \caption{Comparison between our model and the ``long-term'' model. The first row shows a dance stretching motion and the second row shows a kicking motion. ``Long-term'' model generates low-quality and unnatural geometic detials for unseen motions.}
    \label{fig:long_term_compare}
    \includegraphics[width=0.95\linewidth]{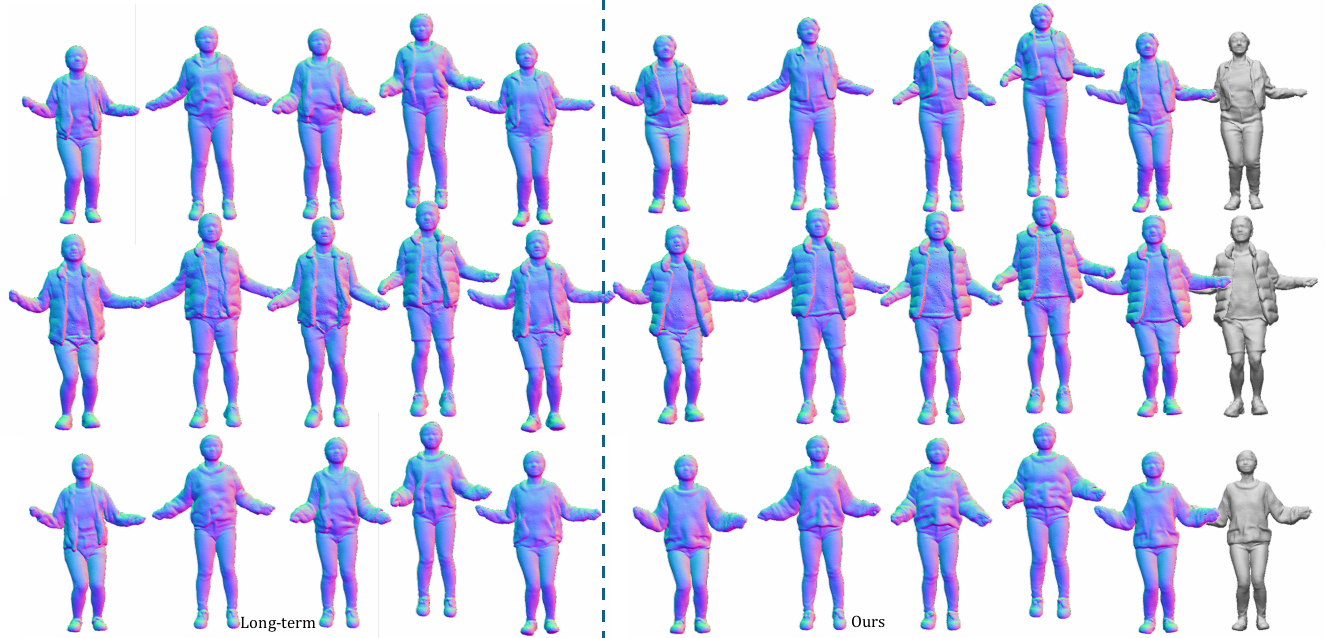}
    \caption{Comparison between our model and the ``long-term'' model. Diverse avatars with a rope-jumping motion are demonstrated. ``Long-term'' model generates low-quality and unnatural geometic detials for unseen motions.}
    \label{fig:long_term_compare2}
\end{figure*}

We provide additional results, including the comparison with the ``w/o augment'' (Fig.~\ref{fig:no_aug_compare}), ``w/o condition'' (Fig.~\ref{fig:id_compare}, Fig.~\ref{fig:id_compare2}), and ``long-term'' (Fig.~\ref{fig:long_term_compare}, Fig~\ref{fig:long_term_compare2}) models, covering different characters and motion types to validate the generalization ability of our method.

\end{document}